\documentclass[aps,prl,twocolumn,superscriptaddress]{revtex4-1}

\usepackage{graphicx}
\usepackage{pgfplots} 
\usepackage{epsfig}
\usepackage{bm}
\usepackage{amsmath}
\usepackage{amssymb}
\usepackage{wasysym}
\usepackage{epstopdf}
\usepackage{color}
\usepackage{xcolor}
\usepackage{epstopdf}
\usepackage{upgreek}
\usepackage{natbib}
\usepackage{arydshln}
\usepackage[T1]{fontenc}
\usepackage[colorlinks=true, bookmarks=true]{hyperref}
\usepackage[normalem]{ulem}
\usepackage{xcolor}
\newcommand\yhsout{\bgroup\markoverwith{\textcolor{blue}{\rule[0.5ex]{2pt}{0.8pt}}}\ULon}
\newcommand\ttsout{\bgroup\markoverwith{\textcolor{red}{\rule[0.5ex]{2pt}{0.8pt}}}\ULon}

\begin{document}

\title{Vibration-induced actuation of droplets on microstructured surfaces}

\author{Dinh-Tuan Phan}
\email{Corresponding author: phan@jhu.edu}
\affiliation{School of Mechanical \& Aerospace Engineering,
	Nanyang Technological University, 
	50 Nanyang Avenue, Singapore 639798}
\affiliation{Current address: Institute for Nanobiotechnology,
Johns Hopkins University, 3400 North Charles Street, Baltimore, Maryland 21218, United States}
\author{Hao Yu}
\affiliation{School of Mechanical \& Aerospace Engineering, 
	Nanyang Technological University, 
	50 Nanyang Avenue, Singapore 639798}
\author{Tuan Tran}
\affiliation{School of Mechanical \& Aerospace Engineering, 
	Nanyang Technological University, 
	50 Nanyang Avenue, Singapore 639798}

\date{\today}

\begin{abstract}
When a liquid droplet impacts a vibrated micro-structured surface with asymmetric topology, the liquids perform a horizontal motion during its bouncing. The moving effect is observed when the liquid is in contact with a low surface energy surface ({\it e.g.} hydrophobic) and over a wide amplitude and frequency range. We propose that liquid droplet's motion direction is driven by a force exerted by the unbalanced vapor flow between the contact of solid and the liquid due to the asymmetric geometry. We observe the levitation and movement dynamics of the droplet impacting on an vibrated micro-structured surface to reveal the processes responsible for the transitional regime between the moving, unmoved and broken droplet as the vibration amplitude and frequency increases. Based on the insight provided by the experiment and on the analysis of the kinetic energy of the droplet, we develop a quantitative model for the dynamic movement and its dependence on the vibration characteristics.
\end{abstract}


\maketitle

When a droplet comes in contact with a solid surface, depending on the surface properties, such as temperature, structure, it can exhibit different impact behaviors.  For examples, droplet can be supported by a thin vapor layer in the Leidenfrost regime~\cite{Biance2003,Tran2012} if the surface is superheated, or the droplet can be bounced~\cite{Richard2000, Richard2002, Jung2008} if the surface is superhydrophobic. With a solid contact surface, droplet can perform self-running due to the chemical energy~\cite{Sumino2005, Ichimura2000}, electrical gradient~\cite{Pollack2000} or surface tension energy~\cite{Daniel2001}; self-propelling~\cite{Linke2006} due to the thermal gradient or bouncing on a wet, inclined surface covered with a thin layer of high viscosity fluid~\cite{Gilet2012}. 
\begin{figure}[t!]
\begin{center}
\includegraphics[width=8.6cm]{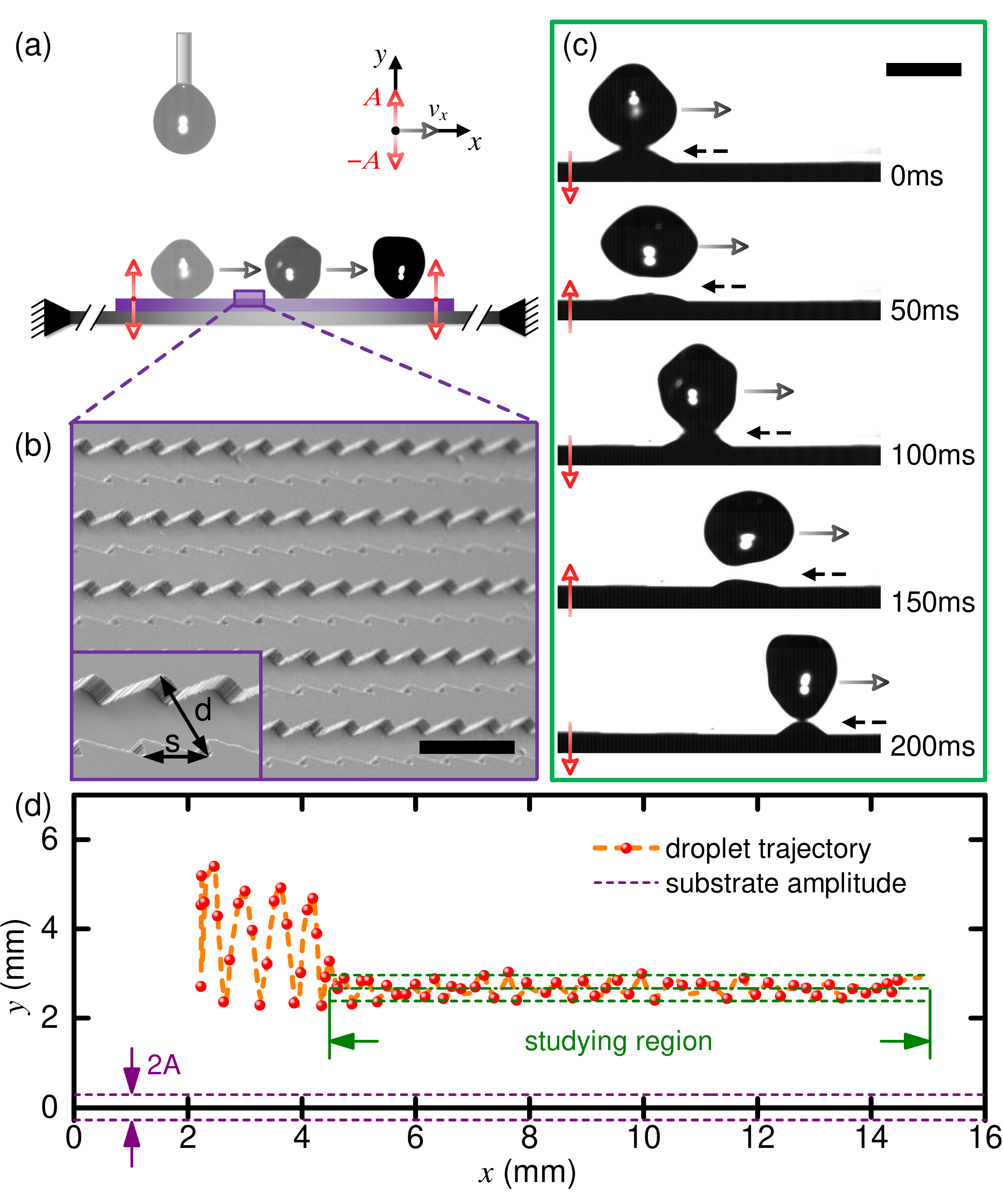}
\caption{
(color online) (a) Schematic 
of the experimental setup (not to scale). 
The vertical red hollow arrows 
indicate the vibration of substrate. 
The horizontal grey hollow arrows 
indicate the droplet moving direction.
(b) SEM micrograph of 
an Silicon substrate with etched parallel microchannels
(scale bar is 100$\,\mu$m). 
Bottom-left inset shows the characteristic 
dimensions of the asymmetric 
in-plane ratchet-like microstructures. 
(c) Representative series of snapshots 
showing actuation of a droplet 
on the oscillating substrate (scale bar is 2$\,$mm). 
The amplitude and frequency are $A$ = 194$\,\mu$m 
and $f$ = 160$\,$Hz, respectively. 
The vertical red hollow arrows 
indicate the instant vibrating direction of the substrate. 
The horizontal grey hollow arrows 
indicate the droplet moving direction.
The horizontal dashed-line arrows
indicate the instant position of the substrate. 
(d) Typical trajectory of a bouncing and moving droplet. The amplitude and frequency are $A$ = 288$\,\mu$m 
and $f$ = 170$\,$Hz, respectively.
Y axis denotes the relative bouncing height, 
which was measured from the center mass of 
droplet to the impact surface. 
X axis denotes the relative horizontal displacement.
\label{fig1}}
\end{center}
\vspace{-0.7cm}
\end{figure}

Traditionally, Leidenfrost state can be used to levitate the droplets caused by a thin vapor layer between the droplet and the heated surface. In this case, droplet can be moved with a minimum force exerting on its due to its nearly frictionless state. Especially, Kruse~\cite{Kruse2015} demonstrated that due to the redirected of vapor flow caused by the inclined angle of mound-like structures, the droplet can be moved in the opposite direction compared to that of conventional ratchet structures~\cite{Linke2006}. 

In the case of a liquid contact surface, droplets shown the walking behaviors on the vibrating fluid bath~\cite{Couder2005,Molacek2013,Wind-Willassen2013} and vibrating rotating fluid bath~\cite{Harris2013}, or dancing~\cite{Vandewalle2006,Terwagne2007}. Gilet~\cite{Gilet2008} demonstrated that the low viscosity oil droplet can be bounced periodically on the high viscosity oil bath when the acceleration of its sinusoidal motion is larger than the resonance frequency of droplet and bath motions. Hence, the droplets are manipulated without any direct contacts between the droplets and the solid interface. It leads to the droplets not being contaminated.

Recently, in the systems of droplet moving on a heated solid surface, such as ratchet surface~\cite{Linke2006} or three-dimensional self-assembled microstructured surfaces~\cite{Kruse2015}, millimeter-scale liquid droplets on substrates presented an effective movement due to an imbalance of viscous forces caused by the vapor layer and can be moved up to 40 cm/s~\cite{Linke2006}. However, these surface require an external continuous supplied temperature and may not be suitable to control its movement characteristics, e.g. velocity, trajectory because its dependence on the Leidenfrost temperature. To the best of our knowledge, droplet impact on an asymmetric structured-surface coupled with a vertical vibration has not been studied yet. 
Such phenomenon has high potential in
manipulating droplet motion without requiring 
thermal or chemical gradients.
Possible applications of such capability include
lab-on-a-chip technologies in which it is desired to manipulate minute amounts
of liquids with minimal invasion to the transported substance,
as well as ease in both chip fabrication processes and controlling methods.

In the present study, we report a simple system for
droplet actuation using vibrating substrates
decorated by asymmetric microstructures. 
We systematically characterize 
the droplet motion including 
the opperating range of the control parameters such as 
the vibration frequency and amplitude. 
We also provide details description of the dynamical behaviors
and proposed a simple analysis to account for the actuation mechanism. 

\begin{figure}[t!]
\begin{center}
\includegraphics[width=9cm]{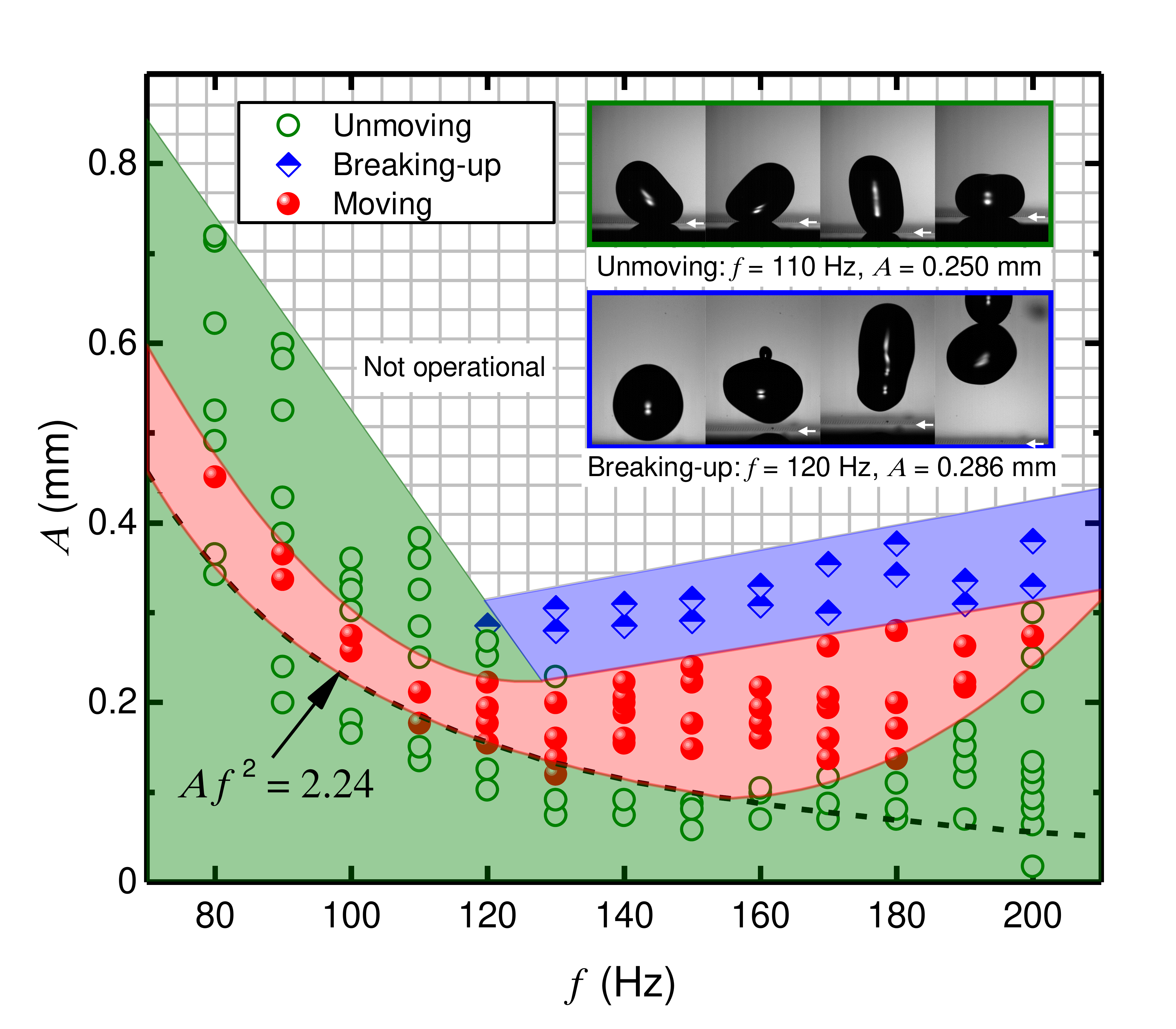}
\caption{
(color online) Phase diagram
for different behaviours of water droplets 
on vibrated substrate. 
Axes: vibration amplitude $A$ and frequency $f$. 
There are three readily separated regimes:
(I) moving regime (red spheres in red block), 
(II) unmoving regime (green open circles in green blocks), and 
(III) breaking-up regime (blue half--open diamonds in blue block). 
Grid region denotes the non--operational experimental conditions.
Insets show representative 
series of snapshots of the unmoving regime and 
the breaking-up regime.
Red arrows indicate the instant position of the substrate.
For representative snapshots
of droplets in the moving regime, 
refer to Fig.~\ref{fig1}.  
The dashed line indicates the transition 
between the moving and the unmoving regimes, 
following the relation 
$Af^2 = 2.24\,{\rm m/s^2}$.
\label{fig2}}
\end{center}
\vspace{-0.7cm}
\end{figure}

Fig.~\ref{fig1}(a) shows the setup of 
droplet actuation experiments. 
Deionized water droplets
of typical diameter $d_0 \approx 1.8\,$mm were released 
from a needle (G30 with an 
outer diameter of 310$\,\mu$m).
The droplets impact the substrate 
at a fixed velocity $v_0 \approx 0.24\,{\rm m/s}$.
The dimension of substrate is 22$\,\times\,$11$\,$mm, 
on which 
the microstructured region of
16.35$\,\times\,$5.15$\,$mm was patterned.  
The substrate was etched to create 
parallel microchannels with wavy walls as shown
in Fig.~\ref{fig1}(b)).
The microchannels have uniform depth of 
50$\,\mu$m
and average width of 
40\,$\mu$m. 
The wavy shape of the microchannel wall
was made asymmetrically in order to 
impose a preferred direction on the air flow 
in the microchannels. 
The substrate was then coated with a thin layer of
commercial hydrophobic solution (Glaco Rain-X)
resulting in a water contact angle of
153$\,^{\circ}$. 
In order to induce vibration of the substrate with controlled 
frequency ($f$) and amplitude ($A$), 
we glued the backside of the substrate to 
the central region of
a horizontal flexible acrylic sheet (1.25\,mm in thickness 
and 208\,mm in diameter).
With this arrangement, it was possible to 
impose vibration to the substrate in the vertical direction only; 
any horizontal movement was eliminated 
as the edge of the acrylic sheet was fixed. 
By using high speed camera (Photron Fastcam SAX-2),
we also confirmed 
that the substrate always remained horizontal 
during vibration, and its motion 
was strictly limited in the vertical direction.
This was to ensure that only the asymmetric microstructures
were responsible for any induced motion of the droplet.
The acrylic sheet's vibration was induced by an
AC speaker (Visaton Speaker BG 20), which was
driven by a function generator 
(LeCroy Wavestation 2012) and 
a power amplifier (Peavey CS4000).
By adjusting the frequency and voltage 
of the applied sinusoidal waveform, 
we were able to control $f$ and $A$ 
of the substrate in the range
$80\,{\rm Hz}\le f \le 200\,{\rm Hz}$
and $100\,{\rm \mu m}\le A \le 400\,{\rm \mu m}$, respectively.

By varying the control parameters $f$ and $A$ 
and observing the droplet behaviors after 
impact on vibrating structured surface, we
identify three characteristic regimes 
(see Fig.~\ref{fig2}). 
(I) Directional moving regime: 
droplets can bounce and move horizontally in  
a certain direction predetermined by 
the microstructure asymmetry on the substrate surface
(see Fig.~\ref{fig1}(b)). 
This regime is the major focus of the present study. 
An important finding of this regime is 
the vibrated vertical acceleration $a_y$ 
defined as $Af^2$ is less than 
the gravitational acceleration ($g$ = 9.8$\,$m/s$^2$), 
except one moving case at $f$ = 200$\,$Hz.
(II) unmoving regime: 
droplets can only bounce on the substrate 
at a fixed position after impact. 
Droplet perform a "spinning" behavior 
with negligible displacement on the structured surface. 
A typical unmoving droplet is exemplified
in inset snapshots with green boundary in Fig.~\ref{fig2}.
(III) breaking-up regime: 
droplets can break into daughter droplets during actuation. 
The droplet is broken 
either at the impact surface or during bounce. 
We observed that for the breaking droplet at the impact surface, 
droplet jumped randomly before breaking.
A typical breaking-up phenomena is presented 
by inset snapshots with blue boundary in Fig.~\ref{fig2}. 

\begin{figure}[t!]
\begin{center}
\includegraphics[width=8.4cm]{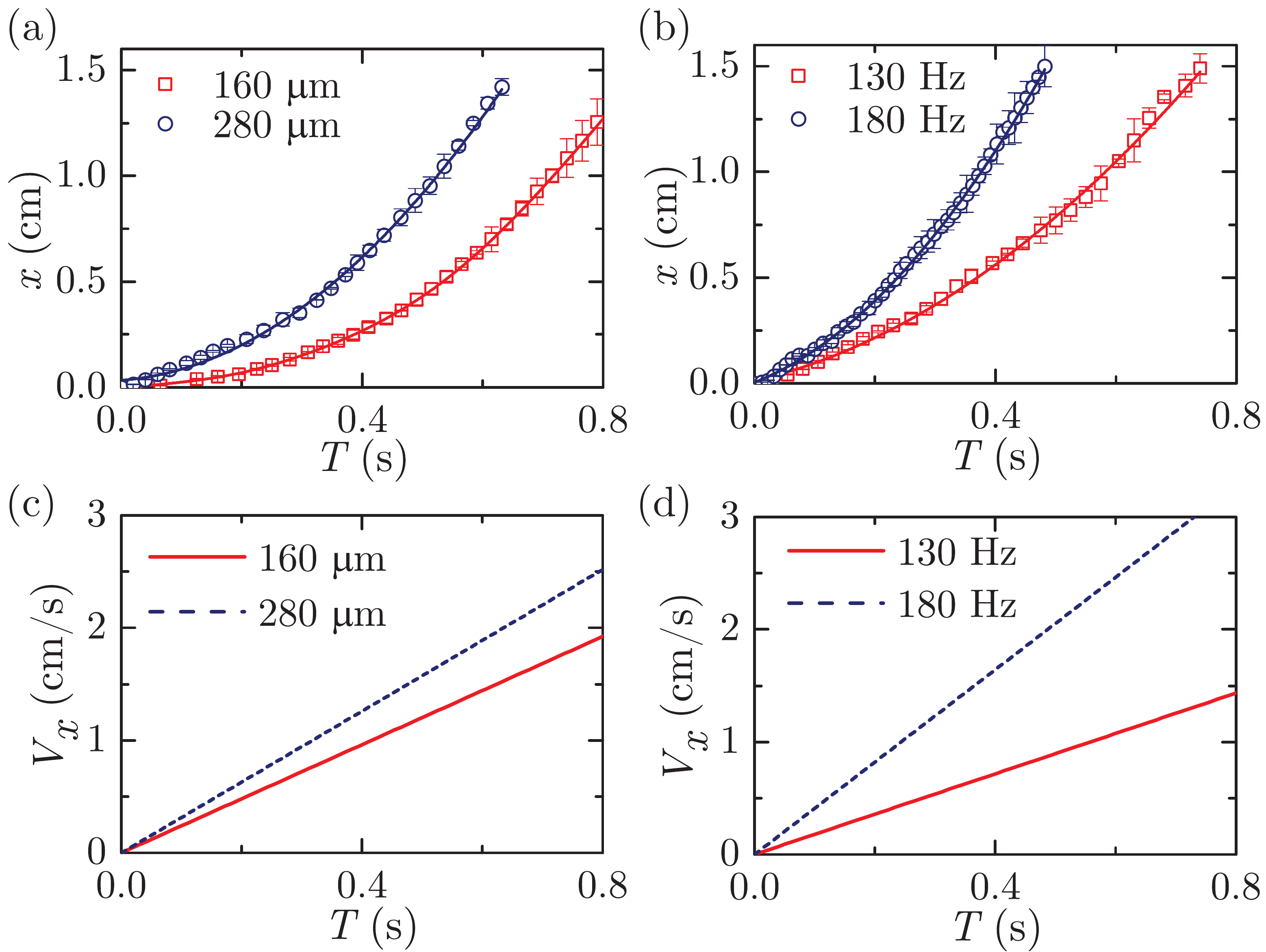}
\caption{
(color online) (a--b) Representative time dependence 
of the horizontal displacement $x$ for 
(a) oscillation amplitudes $A$ = 160 and 280$\,\mu$m
with fixed frequency $f$ = 160$\,$Hz, and 
(b) oscillation frequencies $f$ = 130 and 180$\,$Hz 
with fixed amplitude $A$ = 200$\,\mu$m.
The solid curves represent the best quadratic fitting 
to the experimental data 
with the coefficient of determination $R^2 \geqq 0.95$. 
(c--d) Representative time evolution profiles of 
horizontal moving velocity $V_X$,
calculated as the first derivative of 
the quadratic fitting functions 
corresponding to (a) and (b), respectively. 
\label{fig3}}
\end{center}
\vspace{-0.7cm}
\end{figure}

We propose a movement mechanism of the droplet 
by illustrating its behavior after impact as shown in Fig.~\ref{fig1}(b). 
The illustration is based on the vertical trajectory of centroid droplets and 
be explained experimentally in supplementary material S1.
Basically, when the droplet impacts the vibrated surface, it was supplied a kinetics energy.
Due to the asymmetric microstructured surface, the force transferring into droplet is not in the vertical direction. The patterned microstructure may redirect the vapor flow escaping from the gap between droplet and the impact surface. Because of the ratchet-like sidewalls along the longitudinal direction, this vapor flow cause the unbalance in the surface tension of droplet. It results in an italic force acting on the droplet from the impact location. By continuously supplying the external energy via the vibration, the droplet move in the predefined direction after each bouncing. 
At the first impact of droplet releasing from the height of needle tip, a part of this potential energy was lost due to the impact. From the observation, we noticed that the bouncing heights were decreased after the first impacts. The droplet was supplied continuously with a constant energy, beside the energy lost for the friction, the rest is transferred to the driven force acting on to the droplet.

The horizontal velocity of each running droplet was calculated by first fitting its horizontal displacement with the quadratic function. Next, we take the first derivative of this function to get the instantaneous velocity with respect to time. Then we plotted this instantaneous velocity versus time, as shown in Fig.~\ref{fig3}(c) and (d).
To characterize the droplet directional motion  
on the microstructured surface,
we extract the horizontal displacement 
of the moving droplet by tracing its centroid 
trajectory along the longitudinal direction of 
the asymmetric microchannels. 
The runway of droplet is in the range of 13 -- 15$\,$mm, depending on the initial impact location. We measure the displacements with the time interval of 25 ms.
Representative time evolution profiles of 
horizontal displacement $x$ are shown 
in Fig.~\ref{fig3}(a) and (b). 
Longer horizontal moving distance can be obtained 
with either a larger oscillation amplitude 
(see Fig.~\ref{fig3}(a)) or 
a higher oscillation frequency 
(see Fig.~\ref{fig3}(b)), 
which indicates that more impacting energy 
input to the bouncing droplet 
from the oscillating substrate 
can result in more effective droplet actuation 
in term of moving distance. 
These horizontal displacement profiles 
can then be fitted with quadratic functions, 
represented by the solid curves, 
with the coefficient of determination 
$R^2 \geqq 0.95$. 
By calculating the first derivative 
of the horizontal displacement against time, 
we can therefore obtain  
the time evolution of horizontal velocity $V_x$, 
as exemplified in Fig.~\ref{fig3}(c) and (d).  
Correspondingly, a larger oscillation amplitude 
or a higher oscillation frequency 
can generate higher directional actuation velocity.

\begin{figure}[ht!]
\begin{center}
\includegraphics[width=9cm]{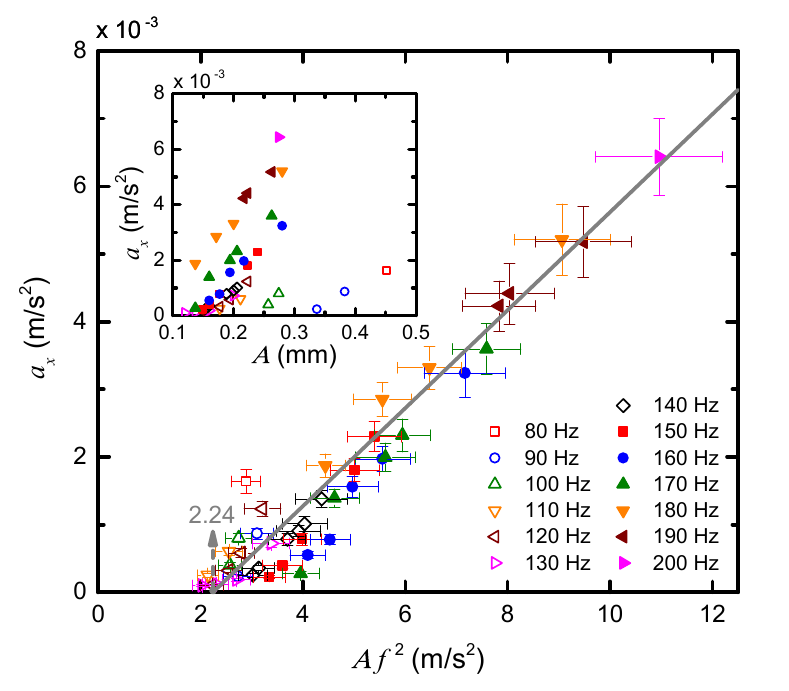}
\caption{
(color online) Horizontal acceleration $a_x$ 
{\it versus} the {\it driving} 
acceleration $a_{d} = Af^2$ in 
the vertical direction.
The grey line represents the best linear fitting 
to the experimental data 
The linear fitting function intersects 
with the $a_{act}$ axis at 2.24$\,$m/s$^2$, 
indicating that the applied oscillation should 
meet the comprehensive condition 
$a_{d}\geqq$ 2.24$\,$m/s$^2$ 
in order to achieve the directional motion 
on the substrate surface with asymmetric microstructures.
\label{fig4}}
\end{center}
\vspace{-0.7cm}
\end{figure}

Note that the horizontal driving force acting 
	on the droplet can be assumed constant 
	if we hypothesise that the impact velocity 
	is roughly constant during all the impacts: 
	constant impact velocity results in constant 
	horizontal momentum. This can be justified 
	if we can show that the typical value for 
	falling velocity of droplet is much smaller 
	than the typical moving velocity of the substrate $Af$. 
We notice that the minimum $Af$ is 0.016$\,$m/s.
We further study the effect of actuation force 
exerted on the droplet from  
the oscillating microstructured surface, 
by taking the second derivative of 
the horizontal displacement $x$ as 
the horizontal acceleration $a_x$. 
To correlate $a_x$ to the perpendicular oscillation, 
we define an effective actuation 
acceleration $a_{act}$, which is  
a function of oscillation amplitude $A$ 
and frequency $f$, expressed by 
$a_{act} = Af^2$. This is the vertical acceleration caused by the vibration determining the lift-off condition when comparing to the gravitational acceleration $g$. If the surface go down with the acceleration larger than $g$, the droplet is "lift-off" from the surface.
When the droplet hit the end of running way, 
we observed that droplet stopped 
at the edge of the microstructure. 
So that we can believe that 
the movement of droplet is caused by 
the asymmetric of the structured surface. Compared to what we tested with the flat surface coated a hydrophobic layer, no directional movement of droplet was observed. The droplet performed the jumping actions as reported previously~\cite{Richard2000, Richard2002}. 
Fig.~\ref{fig4} presents the correlation 
between $a_x$ and $a_{act}$ 
in the full studying ranges of $A$ and $f$.
Results show that a droplet can be actuated to move 
with a horizontal acceleration  
up to 6.5$\,$mm/s$^2$ 
at $f$ of 200$\,$Hz and $A$ of 274$\,\mu$m.
All the experimental data can be well 
fitted by a linear fitting line, 
which intersects with the $a_{act}$ axis 
at 2.2$\,$m/s$^2$.
This intersection indicates that 
a necessary comprehensive actuation condition of 
$a_{act}\geqq$ 2.2$\,$m/s$^2$ should be 
satisfied in order to effectively actuate 
the droplet with a directional motion 
by a perpendicularly oscillating surface.
The collapse of data implies that 
the directional acceleration of droplet 
can be linearly controlled 
by tuning $A$ and $f$ of the oscillating surface 
with asymmetric microstructures.
We also consider the lower transition boundary 
between unmoving and moving regime by proposing the droplet would not be directionally moved if the effective actuation acceleration $a_{act}$ $<$ 2.2 m/s$^2$ corresponding to a negative horizontal acceleration $a_x$. The correlation $Af^2$ = 2.2 m/s$^2$ was then plotted onto the phase diagram in Fig.~\ref{fig2}. 
Interestingly, the two lower transition boundaries created by 
experimental results and by empirical argument 
are almost overlapped.

In conclusion, we 
demostrate that the symmetry breaking of 
microstructures on a vibrating substrate 
effectively causes horizontal acutation of droplets. 
We provide a detailed characterisation of 
the droplet behaviours based on two control parameters: 
the amplitude $A$ and frequency $f$ of the vibrated substrate. 
For the values of control parameters that results in droplet 
actuation, we experimentally found the necessary conditions
for droplet actuation. We proposed that the 
horizontal motion of the droplet is caused by 
viscous stress from the gas flow in the gap between 
microstructures. Although this hypothesis needs further
investigation both numerically and experimentally, 
it effectively demonstrates the feasibility of 
precise control of droplet 
actuations without contact to the base solid surface.
We anticipate that this could be used 
in microfluidics applications~\cite{Liu_2017, Das_2017, Phan_2014apl, Phan_2015jmm, Phan_2015rsc, Phan_2018lc} to control droplet motions 
with minimum contamination and damaging effects to 
the transported agents.

\begin{acknowledgments}
This work was supported by the Nanyang Technological University, Singapore. D.-T. Phan and H. Yu wishes to acknowledge the research fellowship from Nanyang Technological University.

\end{acknowledgments}

\bibliography{prl_tuan}

\begin{thebibliography}{25}%
\makeatletter
\providecommand \@ifxundefined [1]{%
 \@ifx{#1\undefined}
}%
\providecommand \@ifnum [1]{%
 \ifnum #1\expandafter \@firstoftwo
 \else \expandafter \@secondoftwo
 \fi
}%
\providecommand \@ifx [1]{%
 \ifx #1\expandafter \@firstoftwo
 \else \expandafter \@secondoftwo
 \fi
}%
\providecommand \natexlab [1]{#1}%
\providecommand \enquote  [1]{``#1''}%
\providecommand \bibnamefont  [1]{#1}%
\providecommand \bibfnamefont [1]{#1}%
\providecommand \citenamefont [1]{#1}%
\providecommand \href@noop [0]{\@secondoftwo}%
\providecommand \href [0]{\begingroup \@sanitize@url \@href}%
\providecommand \@href[1]{\@@startlink{#1}\@@href}%
\providecommand \@@href[1]{\endgroup#1\@@endlink}%
\providecommand \@sanitize@url [0]{\catcode `\\12\catcode `\$12\catcode
  `\&12\catcode `\#12\catcode `\^12\catcode `\_12\catcode `\%12\relax}%
\providecommand \@@startlink[1]{}%
\providecommand \@@endlink[0]{}%
\providecommand \url  [0]{\begingroup\@sanitize@url \@url }%
\providecommand \@url [1]{\endgroup\@href {#1}{\urlprefix }}%
\providecommand \urlprefix  [0]{URL }%
\providecommand \Eprint [0]{\href }%
\providecommand \doibase [0]{http://dx.doi.org/}%
\providecommand \selectlanguage [0]{\@gobble}%
\providecommand \bibinfo  [0]{\@secondoftwo}%
\providecommand \bibfield  [0]{\@secondoftwo}%
\providecommand \translation [1]{[#1]}%
\providecommand \BibitemOpen [0]{}%
\providecommand \bibitemStop [0]{}%
\providecommand \bibitemNoStop [0]{.\EOS\space}%
\providecommand \EOS [0]{\spacefactor3000\relax}%
\providecommand \BibitemShut  [1]{\csname bibitem#1\endcsname}%
\let\auto@bib@innerbib\@empty
\bibitem [{\citenamefont {Biance}\ \emph {et~al.}(2003)\citenamefont {Biance},
  \citenamefont {Clanet},\ and\ \citenamefont {Quéré}}]{Biance2003}%
  \BibitemOpen
  \bibfield  {author} {\bibinfo {author} {\bibfnamefont {A.-L.}\ \bibnamefont
  {Biance}}, \bibinfo {author} {\bibfnamefont {C.}~\bibnamefont {Clanet}}, \
  and\ \bibinfo {author} {\bibfnamefont {D.}~\bibnamefont {Quéré}},\ }\href
  {\doibase 10.1063/1.1572161} {\bibfield  {journal} {\bibinfo  {journal}
  {Phys. Fluids}\ }\textbf {\bibinfo {volume} {15}},\ \bibinfo {pages} {1632}
  (\bibinfo {year} {2003})}\BibitemShut {NoStop}%
\bibitem [{\citenamefont {Tran}\ \emph {et~al.}(2012)\citenamefont {Tran},
  \citenamefont {Staat}, \citenamefont {Prosperetti}, \citenamefont {Sun},\
  and\ \citenamefont {Lohse}}]{Tran2012}%
  \BibitemOpen
  \bibfield  {author} {\bibinfo {author} {\bibfnamefont {T.}~\bibnamefont
  {Tran}}, \bibinfo {author} {\bibfnamefont {H.~J.~J.}\ \bibnamefont {Staat}},
  \bibinfo {author} {\bibfnamefont {A.}~\bibnamefont {Prosperetti}}, \bibinfo
  {author} {\bibfnamefont {C.}~\bibnamefont {Sun}}, \ and\ \bibinfo {author}
  {\bibfnamefont {D.}~\bibnamefont {Lohse}},\ }\href {\doibase
  10.1103/PhysRevLett.108.036101} {\bibfield  {journal} {\bibinfo  {journal}
  {Phys. Rev. Lett.}\ }\textbf {\bibinfo {volume} {108}},\ \bibinfo {pages} {1}
  (\bibinfo {year} {2012})}\BibitemShut {NoStop}%
\bibitem [{\citenamefont {Richard}\ and\ \citenamefont
  {Qu{\'{e}}r{\'{e}}}(2000)}]{Richard2000}%
  \BibitemOpen
  \bibfield  {author} {\bibinfo {author} {\bibfnamefont {D.}~\bibnamefont
  {Richard}}\ and\ \bibinfo {author} {\bibfnamefont {D.}~\bibnamefont
  {Qu{\'{e}}r{\'{e}}}},\ }\href {\doibase 10.1209/epl/i2000-00547-6} {\bibfield
   {journal} {\bibinfo  {journal} {Europhys. Lett.}\ }\textbf {\bibinfo
  {volume} {50}},\ \bibinfo {pages} {769} (\bibinfo {year} {2000})}\BibitemShut
  {NoStop}%
\bibitem [{\citenamefont {Richard}\ \emph {et~al.}(2002)\citenamefont
  {Richard}, \citenamefont {Clanet},\ and\ \citenamefont
  {Qu{\'{e}}r{\'{e}}}}]{Richard2002}%
  \BibitemOpen
  \bibfield  {author} {\bibinfo {author} {\bibfnamefont {D.}~\bibnamefont
  {Richard}}, \bibinfo {author} {\bibfnamefont {C.}~\bibnamefont {Clanet}}, \
  and\ \bibinfo {author} {\bibfnamefont {D.}~\bibnamefont
  {Qu{\'{e}}r{\'{e}}}},\ }\href {\doibase 10.1038/417811a} {\bibfield
  {journal} {\bibinfo  {journal} {Nature}\ }\textbf {\bibinfo {volume} {417}},\
  \bibinfo {pages} {811} (\bibinfo {year} {2002})}\BibitemShut {NoStop}%
\bibitem [{\citenamefont {Jung}\ and\ \citenamefont
  {Bhushan}(2008)}]{Jung2008}%
  \BibitemOpen
  \bibfield  {author} {\bibinfo {author} {\bibfnamefont {Y.~C.}\ \bibnamefont
  {Jung}}\ and\ \bibinfo {author} {\bibfnamefont {B.}~\bibnamefont {Bhushan}},\
  }\href {\doibase 10.1021/la8003504} {\bibfield  {journal} {\bibinfo
  {journal} {Langmuir}\ }\textbf {\bibinfo {volume} {24}},\ \bibinfo {pages}
  {6262} (\bibinfo {year} {2008})}\BibitemShut {NoStop}%
\bibitem [{\citenamefont {Sumino}\ \emph {et~al.}(2005)\citenamefont {Sumino},
  \citenamefont {Magome}, \citenamefont {Hamada},\ and\ \citenamefont
  {Yoshikawa}}]{Sumino2005}%
  \BibitemOpen
  \bibfield  {author} {\bibinfo {author} {\bibfnamefont {Y.}~\bibnamefont
  {Sumino}}, \bibinfo {author} {\bibfnamefont {N.}~\bibnamefont {Magome}},
  \bibinfo {author} {\bibfnamefont {T.}~\bibnamefont {Hamada}}, \ and\ \bibinfo
  {author} {\bibfnamefont {K.}~\bibnamefont {Yoshikawa}},\ }\href {\doibase
  10.1103/PhysRevLett.94.068301} {\bibfield  {journal} {\bibinfo  {journal}
  {Phys. Rev. Lett.}\ }\textbf {\bibinfo {volume} {94}},\ \bibinfo {pages} {1}
  (\bibinfo {year} {2005})}\BibitemShut {NoStop}%
\bibitem [{\citenamefont {Ichimura}(2000)}]{Ichimura2000}%
  \BibitemOpen
  \bibfield  {author} {\bibinfo {author} {\bibfnamefont {K.}~\bibnamefont
  {Ichimura}},\ }\href {\doibase 10.1126/science.288.5471.1624} {\bibfield
  {journal} {\bibinfo  {journal} {Science (80-. ).}\ }\textbf {\bibinfo
  {volume} {288}},\ \bibinfo {pages} {1624} (\bibinfo {year}
  {2000})}\BibitemShut {NoStop}%
\bibitem [{\citenamefont {Pollack}\ \emph {et~al.}(2000)\citenamefont
  {Pollack}, \citenamefont {Fair},\ and\ \citenamefont
  {Shenderov}}]{Pollack2000}%
  \BibitemOpen
  \bibfield  {author} {\bibinfo {author} {\bibfnamefont {M.~G.}\ \bibnamefont
  {Pollack}}, \bibinfo {author} {\bibfnamefont {R.~B.}\ \bibnamefont {Fair}}, \
  and\ \bibinfo {author} {\bibfnamefont {A.~D.}\ \bibnamefont {Shenderov}},\
  }\href {\doibase 10.1063/1.1308534} {\bibfield  {journal} {\bibinfo
  {journal} {Appl. Phys. Lett.}\ }\textbf {\bibinfo {volume} {77}},\ \bibinfo
  {pages} {1725} (\bibinfo {year} {2000})}\BibitemShut {NoStop}%
\bibitem [{\citenamefont {Daniel}(2001)}]{Daniel2001}%
  \BibitemOpen
  \bibfield  {author} {\bibinfo {author} {\bibfnamefont {S.}~\bibnamefont
  {Daniel}},\ }\href {\doibase 10.1126/science.291.5504.633} {\bibfield
  {journal} {\bibinfo  {journal} {Science (80-. ).}\ }\textbf {\bibinfo
  {volume} {291}},\ \bibinfo {pages} {633} (\bibinfo {year}
  {2001})}\BibitemShut {NoStop}%
\bibitem [{\citenamefont {Linke}\ \emph {et~al.}(2006)\citenamefont {Linke},
  \citenamefont {Alem{\'{a}}n}, \citenamefont {Melling}, \citenamefont
  {Taormina}, \citenamefont {Francis}, \citenamefont {Dow-Hygelund},
  \citenamefont {Narayanan}, \citenamefont {Taylor},\ and\ \citenamefont
  {Stout}}]{Linke2006}%
  \BibitemOpen
  \bibfield  {author} {\bibinfo {author} {\bibfnamefont {H.}~\bibnamefont
  {Linke}}, \bibinfo {author} {\bibfnamefont {B.~J.}\ \bibnamefont
  {Alem{\'{a}}n}}, \bibinfo {author} {\bibfnamefont {L.~D.}\ \bibnamefont
  {Melling}}, \bibinfo {author} {\bibfnamefont {M.~J.}\ \bibnamefont
  {Taormina}}, \bibinfo {author} {\bibfnamefont {M.~J.}\ \bibnamefont
  {Francis}}, \bibinfo {author} {\bibfnamefont {C.~C.}\ \bibnamefont
  {Dow-Hygelund}}, \bibinfo {author} {\bibfnamefont {V.}~\bibnamefont
  {Narayanan}}, \bibinfo {author} {\bibfnamefont {R.~P.}\ \bibnamefont
  {Taylor}}, \ and\ \bibinfo {author} {\bibfnamefont {a.}~\bibnamefont
  {Stout}},\ }\href {\doibase 10.1103/PhysRevLett.96.154502} {\bibfield
  {journal} {\bibinfo  {journal} {Phys. Rev. Lett.}\ }\textbf {\bibinfo
  {volume} {96}},\ \bibinfo {pages} {154502} (\bibinfo {year}
  {2006})}\BibitemShut {NoStop}%
\bibitem [{\citenamefont {Gilet}\ and\ \citenamefont {Bush}(2012)}]{Gilet2012}%
  \BibitemOpen
  \bibfield  {author} {\bibinfo {author} {\bibfnamefont {T.}~\bibnamefont
  {Gilet}}\ and\ \bibinfo {author} {\bibfnamefont {J.~W.~M.}\ \bibnamefont
  {Bush}},\ }\href {\doibase 10.1063/1.4771605} {\bibfield  {journal} {\bibinfo
   {journal} {Phys. Fluids}\ }\textbf {\bibinfo {volume} {24}} (\bibinfo {year}
  {2012}),\ 10.1063/1.4771605}\BibitemShut {NoStop}%
\bibitem [{\citenamefont {Kruse}\ \emph {et~al.}(2015)\citenamefont {Kruse},
  \citenamefont {Somanas}, \citenamefont {Anderson}, \citenamefont {Wilson},
  \citenamefont {Zuhlke}, \citenamefont {Alexander}, \citenamefont {Gogos},\
  and\ \citenamefont {Ndao}}]{Kruse2015}%
  \BibitemOpen
  \bibfield  {author} {\bibinfo {author} {\bibfnamefont {C.}~\bibnamefont
  {Kruse}}, \bibinfo {author} {\bibfnamefont {I.}~\bibnamefont {Somanas}},
  \bibinfo {author} {\bibfnamefont {T.}~\bibnamefont {Anderson}}, \bibinfo
  {author} {\bibfnamefont {C.}~\bibnamefont {Wilson}}, \bibinfo {author}
  {\bibfnamefont {C.}~\bibnamefont {Zuhlke}}, \bibinfo {author} {\bibfnamefont
  {D.}~\bibnamefont {Alexander}}, \bibinfo {author} {\bibfnamefont
  {G.}~\bibnamefont {Gogos}}, \ and\ \bibinfo {author} {\bibfnamefont
  {S.}~\bibnamefont {Ndao}},\ }\href {\doibase 10.1007/s10404-014-1540-6}
  {\bibfield  {journal} {\bibinfo  {journal} {Microfluid. Nanofluidics}\
  }\textbf {\bibinfo {volume} {18}},\ \bibinfo {pages} {1417} (\bibinfo {year}
  {2015})}\BibitemShut {NoStop}%
\bibitem [{\citenamefont {Couder}\ \emph {et~al.}(2005)\citenamefont {Couder},
  \citenamefont {Fort}, \citenamefont {Gautier},\ and\ \citenamefont
  {Boudaoud}}]{Couder2005}%
  \BibitemOpen
  \bibfield  {author} {\bibinfo {author} {\bibfnamefont {Y.}~\bibnamefont
  {Couder}}, \bibinfo {author} {\bibfnamefont {E.}~\bibnamefont {Fort}},
  \bibinfo {author} {\bibfnamefont {C.-H.}\ \bibnamefont {Gautier}}, \ and\
  \bibinfo {author} {\bibfnamefont {a.}~\bibnamefont {Boudaoud}},\ }\href
  {\doibase 10.1103/PhysRevLett.94.177801} {\bibfield  {journal} {\bibinfo
  {journal} {Phys. Rev. Lett.}\ }\textbf {\bibinfo {volume} {94}},\ \bibinfo
  {pages} {177801} (\bibinfo {year} {2005})}\BibitemShut {NoStop}%
\bibitem [{\citenamefont {Mol{\'{a}}{\v{c}}ek}\ and\ \citenamefont
  {Bush}(2013)}]{Molacek2013}%
  \BibitemOpen
  \bibfield  {author} {\bibinfo {author} {\bibfnamefont {J.}~\bibnamefont
  {Mol{\'{a}}{\v{c}}ek}}\ and\ \bibinfo {author} {\bibfnamefont {J.~W.~M.}\
  \bibnamefont {Bush}},\ }\href {\doibase 10.1017/jfm.2013.280} {\bibfield
  {journal} {\bibinfo  {journal} {J. Fluid Mech.}\ }\textbf {\bibinfo {volume}
  {727}},\ \bibinfo {pages} {612} (\bibinfo {year} {2013})}\BibitemShut
  {NoStop}%
\bibitem [{\citenamefont {Wind-Willassen}\ \emph {et~al.}(2013)\citenamefont
  {Wind-Willassen}, \citenamefont {Mol{\'{a}}{\v{c}}ek}, \citenamefont
  {Harris},\ and\ \citenamefont {Bush}}]{Wind-Willassen2013}%
  \BibitemOpen
  \bibfield  {author} {\bibinfo {author} {\bibfnamefont {{\O}.}~\bibnamefont
  {Wind-Willassen}}, \bibinfo {author} {\bibfnamefont {J.}~\bibnamefont
  {Mol{\'{a}}{\v{c}}ek}}, \bibinfo {author} {\bibfnamefont {D.~M.}\
  \bibnamefont {Harris}}, \ and\ \bibinfo {author} {\bibfnamefont {J.~W.~M.}\
  \bibnamefont {Bush}},\ }\href {\doibase 10.1063/1.4817612} {\bibfield
  {journal} {\bibinfo  {journal} {Phys. Fluids}\ }\textbf {\bibinfo {volume}
  {25}} (\bibinfo {year} {2013}),\ 10.1063/1.4817612}\BibitemShut {NoStop}%
\bibitem [{\citenamefont {Harris}\ and\ \citenamefont
  {Bush}(2013)}]{Harris2013}%
  \BibitemOpen
  \bibfield  {author} {\bibinfo {author} {\bibfnamefont {D.~M.}\ \bibnamefont
  {Harris}}\ and\ \bibinfo {author} {\bibfnamefont {J.~W.~M.}\ \bibnamefont
  {Bush}},\ }\href {\doibase 10.1017/jfm.2013.627} {\bibfield  {journal}
  {\bibinfo  {journal} {J. Fluid Mech.}\ }\textbf {\bibinfo {volume} {739}},\
  \bibinfo {pages} {444} (\bibinfo {year} {2013})}\BibitemShut {NoStop}%
\bibitem [{\citenamefont {Vandewalle}\ \emph {et~al.}(2006)\citenamefont
  {Vandewalle}, \citenamefont {Terwagne}, \citenamefont {Mulleners},
  \citenamefont {Gilet},\ and\ \citenamefont {Dorbolo}}]{Vandewalle2006}%
  \BibitemOpen
  \bibfield  {author} {\bibinfo {author} {\bibfnamefont {N.}~\bibnamefont
  {Vandewalle}}, \bibinfo {author} {\bibfnamefont {D.}~\bibnamefont
  {Terwagne}}, \bibinfo {author} {\bibfnamefont {K.}~\bibnamefont {Mulleners}},
  \bibinfo {author} {\bibfnamefont {T.}~\bibnamefont {Gilet}}, \ and\ \bibinfo
  {author} {\bibfnamefont {S.}~\bibnamefont {Dorbolo}},\ }\href {\doibase
  10.1063/1.2335905} {\bibfield  {journal} {\bibinfo  {journal} {Phys. Fluids}\
  }\textbf {\bibinfo {volume} {18}},\ \bibinfo {pages} {177801} (\bibinfo
  {year} {2006})}\BibitemShut {NoStop}%
\bibitem [{\citenamefont {Terwagne}\ \emph {et~al.}(2007)\citenamefont
  {Terwagne}, \citenamefont {Vandewalle},\ and\ \citenamefont
  {Dorbolo}}]{Terwagne2007}%
  \BibitemOpen
  \bibfield  {author} {\bibinfo {author} {\bibfnamefont {D.}~\bibnamefont
  {Terwagne}}, \bibinfo {author} {\bibfnamefont {N.}~\bibnamefont
  {Vandewalle}}, \ and\ \bibinfo {author} {\bibfnamefont {S.}~\bibnamefont
  {Dorbolo}},\ }\href {\doibase 10.1103/PhysRevE.76.056311} {\bibfield
  {journal} {\bibinfo  {journal} {Phys. Rev. E - Stat. Nonlinear, Soft Matter
  Phys.}\ }\textbf {\bibinfo {volume} {76}},\ \bibinfo {pages} {1} (\bibinfo
  {year} {2007})},\ \Eprint {http://arxiv.org/abs/0705.4400} {arXiv:0705.4400}
  \BibitemShut {NoStop}%
\bibitem [{\citenamefont {Gilet}\ \emph {et~al.}(2008)\citenamefont {Gilet},
  \citenamefont {Terwagne}, \citenamefont {Vandewalle},\ and\ \citenamefont
  {Dorbolo}}]{Gilet2008}%
  \BibitemOpen
  \bibfield  {author} {\bibinfo {author} {\bibfnamefont {T.}~\bibnamefont
  {Gilet}}, \bibinfo {author} {\bibfnamefont {D.}~\bibnamefont {Terwagne}},
  \bibinfo {author} {\bibfnamefont {N.}~\bibnamefont {Vandewalle}}, \ and\
  \bibinfo {author} {\bibfnamefont {S.}~\bibnamefont {Dorbolo}},\ }\href
  {\doibase 10.1103/PhysRevLett.100.167802} {\bibfield  {journal} {\bibinfo
  {journal} {Phys. Rev. Lett.}\ }\textbf {\bibinfo {volume} {100}},\ \bibinfo
  {pages} {1} (\bibinfo {year} {2008})}\BibitemShut {NoStop}%
\bibitem [{\citenamefont {Liu}\ \emph {et~al.}(2017)\citenamefont {Liu},
  \citenamefont {Xu}, \citenamefont {An}, \citenamefont {Phan}, \citenamefont
  {Hashimoto},\ and\ \citenamefont {Lew}}]{Liu_2017}%
  \BibitemOpen
  \bibfield  {author} {\bibinfo {author} {\bibfnamefont {N.}~\bibnamefont
  {Liu}}, \bibinfo {author} {\bibfnamefont {J.}~\bibnamefont {Xu}}, \bibinfo
  {author} {\bibfnamefont {H.-J.}\ \bibnamefont {An}}, \bibinfo {author}
  {\bibfnamefont {D.-T.}\ \bibnamefont {Phan}}, \bibinfo {author}
  {\bibfnamefont {M.}~\bibnamefont {Hashimoto}}, \ and\ \bibinfo {author}
  {\bibfnamefont {W.~S.}\ \bibnamefont {Lew}},\ }\href {\doibase
  10.1088/1361-6439/aa82ce} {\bibfield  {journal} {\bibinfo  {journal} {Journal
  of Micromechanics and Microengineering}\ }\textbf {\bibinfo {volume} {27}},\
  \bibinfo {pages} {104001} (\bibinfo {year} {2017})}\BibitemShut {NoStop}%
\bibitem [{\citenamefont {Das}\ \emph {et~al.}(2017)\citenamefont {Das},
  \citenamefont {Phan}, \citenamefont {Zhao}, \citenamefont {Kang},
  \citenamefont {Chan},\ and\ \citenamefont {Yang}}]{Das_2017}%
  \BibitemOpen
  \bibfield  {author} {\bibinfo {author} {\bibfnamefont {D.}~\bibnamefont
  {Das}}, \bibinfo {author} {\bibfnamefont {D.-T.}\ \bibnamefont {Phan}},
  \bibinfo {author} {\bibfnamefont {Y.}~\bibnamefont {Zhao}}, \bibinfo {author}
  {\bibfnamefont {Y.}~\bibnamefont {Kang}}, \bibinfo {author} {\bibfnamefont
  {V.}~\bibnamefont {Chan}}, \ and\ \bibinfo {author} {\bibfnamefont
  {C.}~\bibnamefont {Yang}},\ }\href {\doibase 10.1002/elps.201600477}
  {\bibfield  {journal} {\bibinfo  {journal} {Electrophoresis}\ }\textbf
  {\bibinfo {volume} {38}},\ \bibinfo {pages} {645} (\bibinfo {year}
  {2017})}\BibitemShut {NoStop}%
\bibitem [{\citenamefont {Phan}\ and\ \citenamefont
  {Nguyen}(2014)}]{Phan_2014apl}%
  \BibitemOpen
  \bibfield  {author} {\bibinfo {author} {\bibfnamefont {D.-T.}\ \bibnamefont
  {Phan}}\ and\ \bibinfo {author} {\bibfnamefont {N.-T.}\ \bibnamefont
  {Nguyen}},\ }\href {\doibase 10.1063/1.4866970} {\bibfield  {journal}
  {\bibinfo  {journal} {Applied Physics Letters}\ }\textbf {\bibinfo {volume}
  {104}},\ \bibinfo {pages} {084104} (\bibinfo {year} {2014})}\BibitemShut
  {NoStop}%
\bibitem [{\citenamefont {Phan}\ \emph
  {et~al.}(2015{\natexlab{a}})\citenamefont {Phan}, \citenamefont {Yang},\ and\
  \citenamefont {Nguyen}}]{Phan_2015jmm}%
  \BibitemOpen
  \bibfield  {author} {\bibinfo {author} {\bibfnamefont {D.-T.}\ \bibnamefont
  {Phan}}, \bibinfo {author} {\bibfnamefont {C.}~\bibnamefont {Yang}}, \ and\
  \bibinfo {author} {\bibfnamefont {N.-T.}\ \bibnamefont {Nguyen}},\ }\href
  {\doibase 10.1088/0960-1317/25/11/115019} {\bibfield  {journal} {\bibinfo
  {journal} {Journal of Micromechanics and Microengineering}\ }\textbf
  {\bibinfo {volume} {25}},\ \bibinfo {pages} {115019} (\bibinfo {year}
  {2015}{\natexlab{a}})}\BibitemShut {NoStop}%
\bibitem [{\citenamefont {Phan}\ \emph
  {et~al.}(2015{\natexlab{b}})\citenamefont {Phan}, \citenamefont {Chun},\ and\
  \citenamefont {Nguyen}}]{Phan_2015rsc}%
  \BibitemOpen
  \bibfield  {author} {\bibinfo {author} {\bibfnamefont {D.-T.}\ \bibnamefont
  {Phan}}, \bibinfo {author} {\bibfnamefont {Y.}~\bibnamefont {Chun}}, \ and\
  \bibinfo {author} {\bibfnamefont {N.-T.}\ \bibnamefont {Nguyen}},\ }\href
  {\doibase 10.1039/C5RA07491F} {\bibfield  {journal} {\bibinfo  {journal} {RSC
  Adv.}\ }\textbf {\bibinfo {volume} {5}},\ \bibinfo {pages} {44336} (\bibinfo
  {year} {2015}{\natexlab{b}})}\BibitemShut {NoStop}%
\bibitem [{\citenamefont {Phan}\ \emph {et~al.}(2018)\citenamefont {Phan},
  \citenamefont {Jin}, \citenamefont {Wustoni},\ and\ \citenamefont
  {Chen}}]{Phan_2018lc}%
  \BibitemOpen
  \bibfield  {author} {\bibinfo {author} {\bibfnamefont {D.-T.}\ \bibnamefont
  {Phan}}, \bibinfo {author} {\bibfnamefont {L.}~\bibnamefont {Jin}}, \bibinfo
  {author} {\bibfnamefont {S.}~\bibnamefont {Wustoni}}, \ and\ \bibinfo
  {author} {\bibfnamefont {C.-H.}\ \bibnamefont {Chen}},\ }\href {\doibase
  10.1039/C7LC01066D} {\bibfield  {journal} {\bibinfo  {journal} {Lab Chip}\
  }\textbf {\bibinfo {volume} {18}},\ \bibinfo {pages} {574} (\bibinfo {year}
  {2018})}\BibitemShut {NoStop}%
\end{thebibliography}%

\end{document}